\documentclass[fleqn,usenatbib]{mnras}

\usepackage{newtxtext,newtxmath}

\usepackage[T1]{fontenc}

\DeclareRobustCommand{\VAN}[3]{#2}
\let\VANthebibliography\thebibliography
\def\thebibliography{\DeclareRobustCommand{\VAN}[3]{##3}\VANthebibliography}

\usepackage{graphicx}	
\usepackage{amsmath}	

\title[Interstellar Objects in the Oort Cloud]{Interstellar Objects Outnumber Solar System Objects in the Oort Cloud}

\author[A. Siraj \& A. Loeb]{
A. Siraj$^{1}$\thanks{E-mail: amir.siraj@cfa.harvard.edu}
and A. Loeb $^{1}$\thanks{E-mail: aloeb@cfa.harvard.edu}
\\
$^{1}$Department of Astronomy, Harvard University, 60 Garden Street, Cambridge, MA 02138, USA\\
}

\date{Accepted XXX. Received YYY; in original form ZZZ}

\pubyear{2021}

\begin{document}
\label{firstpage}
\pagerange{\pageref{firstpage}--\pageref{lastpage}}
\maketitle

\begin{abstract}
Here, we show that the detection of Borisov implies that interstellar objects outnumber Solar system objects in the Oort cloud, whereas the reverse is true near the Sun due to the stronger gravitational focusing of bound objects. This hypothesis can be tested with stellar occultation surveys of the Oort cloud. Furthermore, we demonstrate that $\sim 1 \%$ of carbon and oxygen in the Milky Way Galaxy may be locked in interstellar objects, implying a heavy element budget for interstellar objects comparable to the heavy element budget of the minimum mass Solar nebula model. There is still considerable uncertainty regarding the size distribution of the interstellar objects.
\end{abstract}

\begin{keywords}
Oort cloud -- asteroids -- meteors
\end{keywords}



\section{Introduction}

The Oort cloud is the Solar System's reservoir of long-period comets \citep{2019AJ....157..181V}. Observations of the first confirmed interstellar comet in the Solar System, Borisov, revealed an object that bore physical resemblance to a long-period comet, yet travelling on a hyperbolic orbit \citep{2020NatAs...4...53G, 2020NatAs...4..867B}. The detection of Borisov, the first confirmed interstellar comet with a known composition \citep{2020NatAs...4...53G, 2020NatAs...4..867B}, implied a number density far from the Sun of $n_B \sim ((4\pi/3) \cdot (3 \mathrm{\; AU})^3)^{-1} = 9 \times 10^{-3} \; \mathrm{AU^{-3}}$, if a single Borisov-like object typically exists within 3 AU of the Sun. This order-of-magnitude calibration mirrors the reasoning used in deriving the abundance of `Oumuamua-like objects \citep{2018ApJ...855L..10D}, and just like the abundance implied by `Oumuamua, Poisson uncertainties of several orders of magnitude dominate the error budget for Borisov-like objects (see Section \ref{ra}). The detection of Borisov is consistent with the number density implied by  `Oumuamua \citep{2020ApJ...888L..23J}.

Here, we study implications of the observational constraint on interstellar object abundance set by Borisov. In Section \ref{ra}, we investigate the relative abundances of interstellar and non-interstellar objects in the Oort cloud, and in Section \ref{heb}, we explore the fraction of heavy metals locked in interstellar objects. 

\section{Relative abundances}
\label{ra}

\begin{figure}
 \centering
\includegraphics[width=1\linewidth]{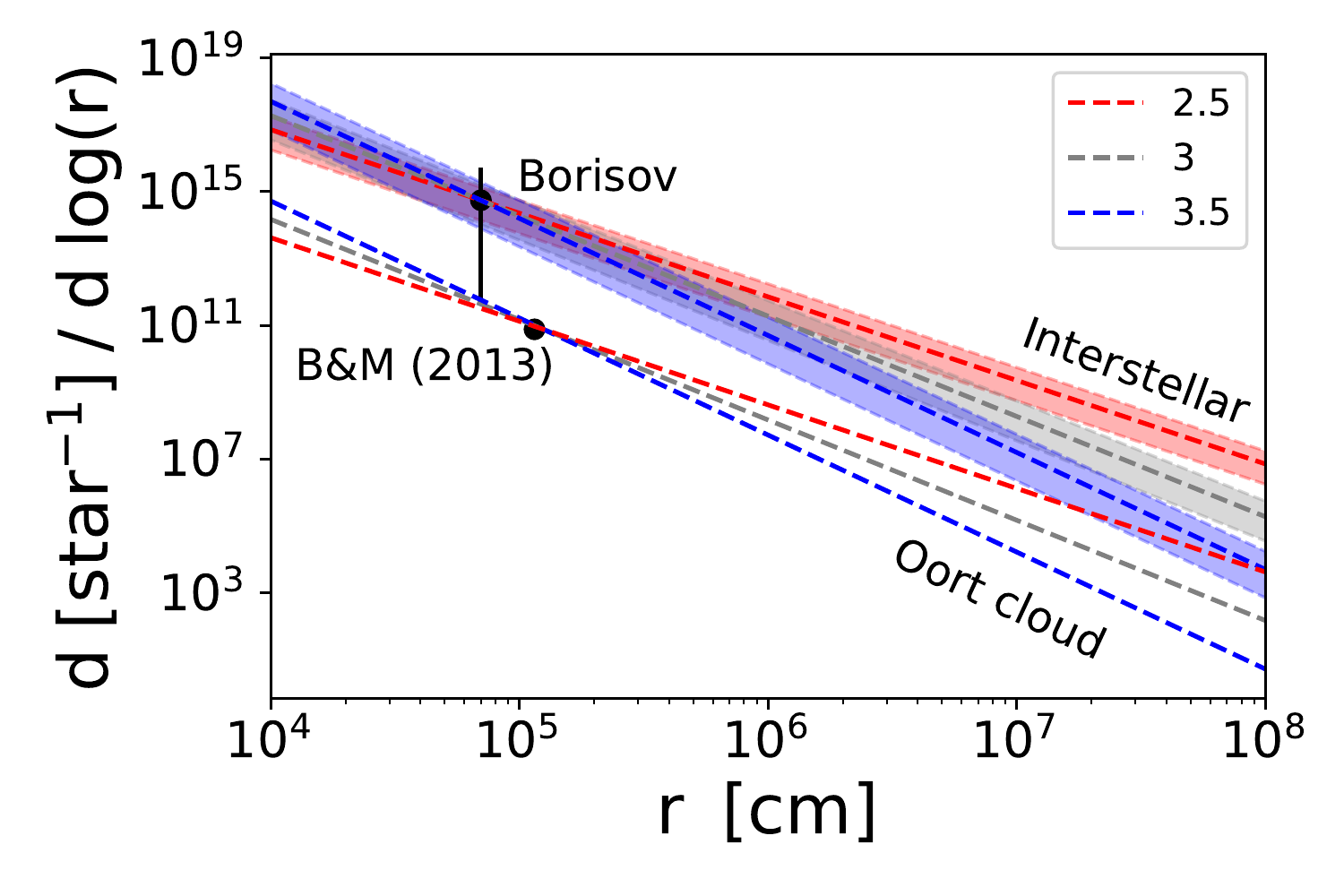}

\caption{Comparison of the relative abundance per star of bound Oort cloud objects, as implied by the observed rate of long-period comets \citep{2013Icar..225...40B}, and interstellar objects, as implied by the detection of Borisov \citep{2020ApJ...888L..23J}, with a differential size distribution for power-law index, $q$, values of 2.5, 3, and 3.5, displayed for reference. The error bar indicates the $3\sigma$ Poisson error bars for the implication of a singular interstellar object detection on the abundance. The shaded band correspond to the plausible range of nucleus radii for Borisov, given the central value for Borisov's abundance. The error bounds on the abundance of bound Oort cloud objects are not resolvable on this plot.}
\label{fig:borsiov_vs_oort}
\end{figure}

The local stellar density of $n_{\star} \simeq 0.14 \; \mathrm{pc^{-3}}$ implies a total number of $N_B \equiv (n_B / n_{\star}) \sim 6 \times 10^{14}$ Borisov-like objects per star \citep{2020ApJ...899L..24S}. The $3 \sigma$ or $99.9 \%$ Poisson confidence interval \citep{crow1959confidence} for the background distribution of interstellar objects implied by the detection of Borisov corresponds to a number density of $n_B \simeq 9 \times 10^{-3^{+0.98}_{-3}} \; \mathrm{AU^{-3}}$ and abundance of $N_B \simeq 6 \times 10^{14^{+0.98}_{-3}}$ Borisov-size objects per star. For a power-law differential size distribution of the form, $dN/dR \propto R^{-q}$, the values of $q$ that apply to comets are in the range $2.5 - 3.5$, with the central value of $q = 3$ representing a scale-free power-law \citep{2012MNRAS.423.1674F}. Additionally, the observational constraints on the radius of Borisov's nucleus \citep{2020ApJ...888L..23J} set a lower limit of $\sim 0.4 \mathrm{\; km}$ and an upper limit of $\sim 1 \mathrm{\; km}$, implying $\sim 0.3 \mathrm{\; km}$ of uncertainty about the central value of $r_B = 0.7 \mathrm{\; km}$. There are $N_{OC} \simeq 7.6 \pm 3.3 \times 10^{10}$ bound Oort cloud objects with diameter $\sim 2.3 \mathrm{\; km}$, inferred from observations of long-period comets \citep{2013Icar..225...40B, 2017A&A...598A.110R}. Figure \ref{fig:borsiov_vs_oort} displays the abundances of Borisov-like unbound interstellar objects and bound Oort cloud objects.

\begin{figure}
 \centering
\includegraphics[width=1\linewidth]{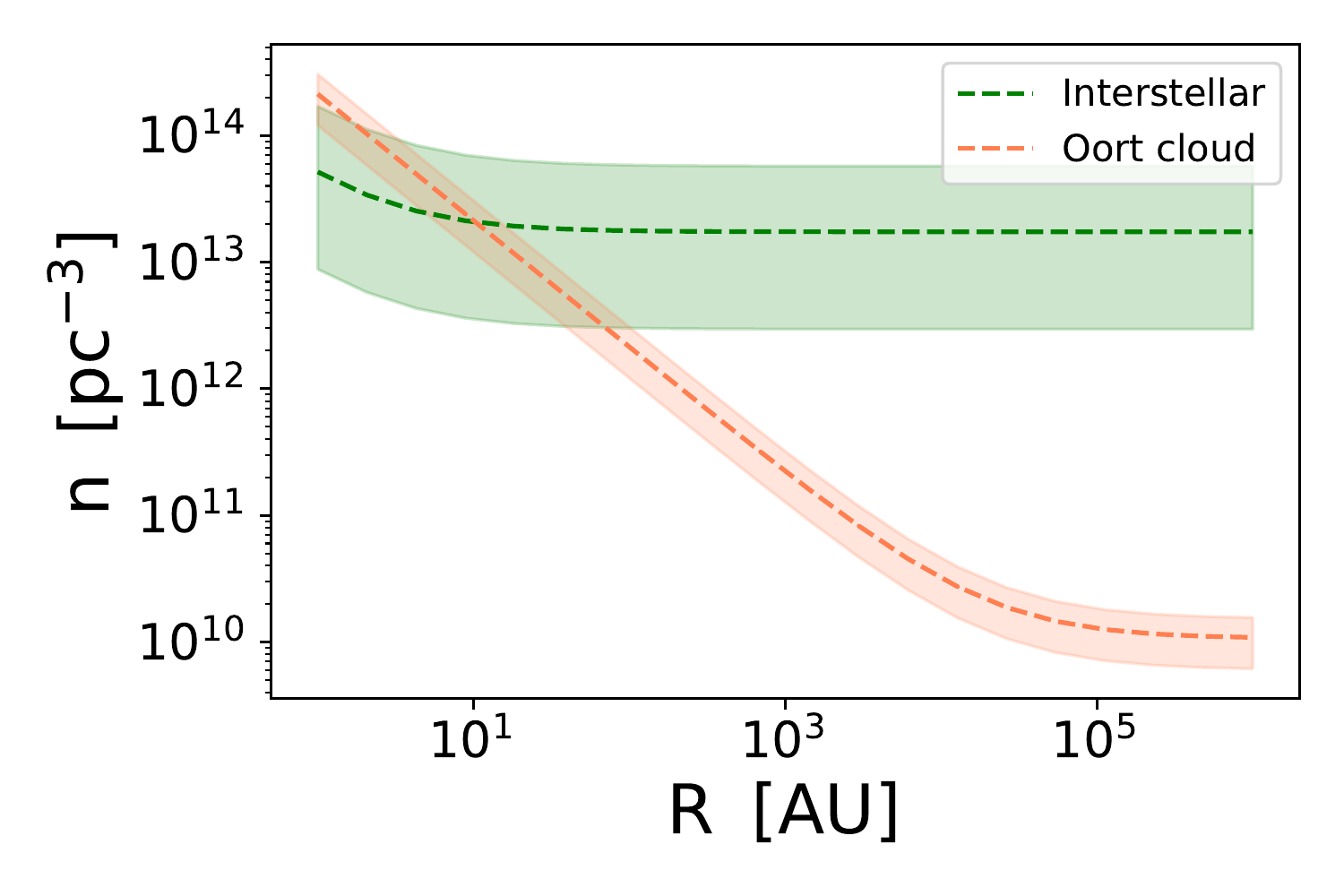}

\caption{Comparison of the local number density, $n$, of bound Oort cloud objects and interstellar objects as a function of distance, $R$, from the Sun due to gravitational focusing, for objects with diameters of $\sim 2.3 \mathrm{\; km}$ and adopting an interstellar object velocity dispersion at infinity of $\sim 30 \mathrm{\; km \; s^{-1}}$ \citep{2020ApJ...903L..20S} and a typical bound Oort cloud object semimajor axis of $\sim 2 \times 10^4 \mathrm{\; AU}$ \citep{2015SSRv..197..191D}. The bands indicate $1\sigma$ Poisson errors for the abundances of interstellar objects and bound Oort cloud objects.}
\label{fig:grav_focusing_vs_oc}
\end{figure}

Adopting the central values of $r_B = 0.7 \mathrm{\; km}$ and $q = 3$, the number density of interstellar objects far from the Sun exceeds significantly that of similarly sized bound Oort cloud objects. The number density of interstellar objects far from the Sun, as implied by Borisov, exceeds the observed number density of bound Oort cloud objects by more than three orders of magnitude, comparable to the $3 \sigma$ Poisson lower bound. Gravitational focusing allows for the comparison of the two populations as a function of distance from the Sun, via the number density scaling factor of $(1 + v_r^2/v_{\infty}^2)$, where $v_R = \sqrt{2GM_{\odot} / R}$ is the orbital speed at distance $R$ from the Sun and $v_{\infty}$ is the characteristic speed at infinity for objects in the population. Given the local stellar velocity dispersion, interstellar objects typically have speeds of $v_{\infty} \sim 30 \mathrm{\; km \; s^{-1}}$ \citep{2020ApJ...903L..20S}, hence experience negligible gravitational focusing from the Sun outside of the Earth's orbit. In contrast, bound Oort cloud objects have typical semimajor axes of $\sim 2 \times 10^4 \mathrm{\; AU}$ \citep{2015SSRv..197..191D}, implying that gravitational focusing significantly affects the number density of bound Oort cloud objects as a function of distance from the Sun outside of the Earth's orbit. This is conditional on the conservative assumption that bound Oort cloud objects fill isotropic orbits, characteristic of trapped origins, as opposed to radially biased orbits, characteristic of scattered origins, which would enhance further the inner Solar system number density of bound Oort cloud objects. Figure \ref{fig:grav_focusing_vs_oc} shows that while interstellar objects are more numerous than bound Oort cloud objects in the Oort cloud ($R \sim 10^5 \mathrm{\; AU}$), bound Oort cloud objects are the dominant population in the inner Solar system, consistent with observations. Present observations of comets are confined to the vicinity of the Earth, a region of the Solar system where the characteristic orbital speed is $\sim 30 \mathrm{\; km \; s^{-1}}$. As a result, bound long-period comets, which have a low velocity dispersion of $\sim 0.3 \mathrm{\; km \; s^{-1}}$ at the outer envelope of the Oort cloud, experience an enhancement by gravitational focusing in their density of order $\sim 10^4$, whereas interstellar objects, which have a higher background velocity dispersion of $\sim 30 \mathrm{\; km \; s^{-1}}$, only experience an enhancement in density of order unity relative to their abundance far from the Sun. Given that the number density of interstellar objects may be $\sim 10^3$ larger than that of bound Oort cloud objects far from the Sun, the Oort cloud objects may be still a factor of $\sim 10$ more abundant than interstellar objects in the inner Solar system, due to the unequal influence of gravitational focusing on the two populations. The fact that interstellar objects outnumber Oort cloud objects per star is consistent with the Oort cloud having lost most of its initial mass. However, the degree to which interstellar objects outnumber Oort cloud objects is still very uncertain. Stellar occultation surveys of the Oort cloud will be capable of confirming the results presented here, by differentiating between the two populations through speed relative to the Sun \citep{2007AJ....134.1596N, 2020ApJ...891L...3S} . 

\begin{figure}[hptb]
 \centering
\includegraphics[width=1\linewidth]{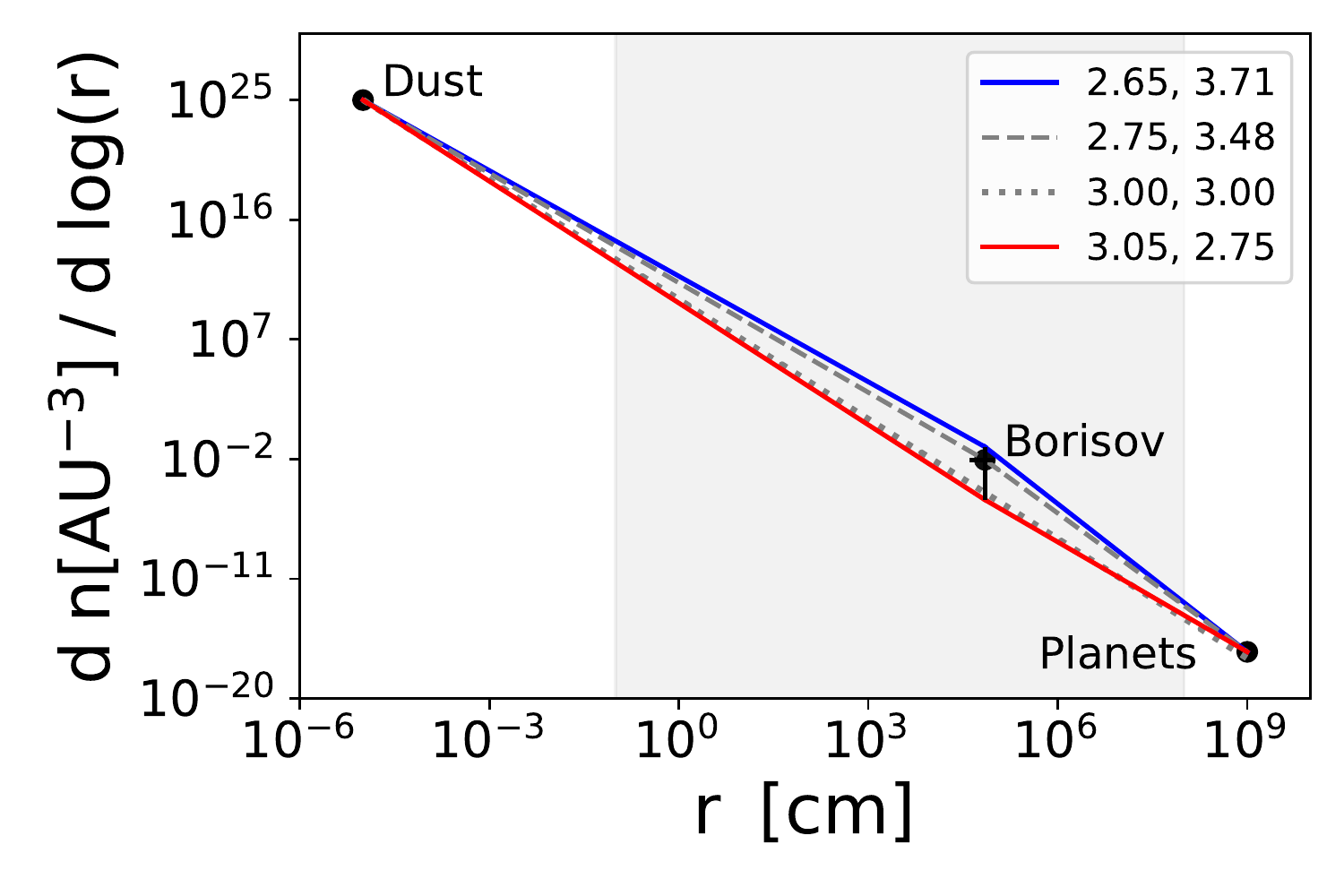}

\caption{Possible broken power-law size distributions for interstellar objects, anchored by the abundances of interstellar dust and rogue planets, and constrained by the implied abundance of Borisov-size interstellar objects with the corresponding $3\sigma$ Poisson errors, and values of $q$ for the two legs of the power-law shown for reference. The blue and red lines correspond to the $3\sigma$ upper and lower bounds on the abundance of Borisov-size objects, respectively, while the gray dashed line intersects the implied abundance of Borisov size-objects. The dotted gray line indicates a $q = 3$ fit, the scale-free power-law size distribution with no break. The gray shaded region indicates the size range which we integrate over in our analysis (see Section \ref{heb} for an explanation for the bounds)}.
\label{fig:borisov_size_dists}
\end{figure}

\section{Heavy element budget}
\label{heb}

Next, we explore the implications of the abundance of interstellar objects on the overall mass fraction of carbon and oxygen they account for. Carbon and oxygen combined dominate the heavy metals budget, representing $\sim 0.82 \%$ of baryons by mass out of the $\sim 1.34 \%$ of baryons by mass in elements heavier than H and He, given photospheric solar composition \citep{2009ARA&A..47..481A}. Interstellar objects like Borisov are primarily composed of carbon and oxygen by mass, in the form of ices, and detection of interstellar dust and of rogue planets can be used to calibrate the overall size distribution, from interstellar micrometeoroids to dwarf planets. 

In particular, interstellar dust grains with a radius of $\sim 10^{-5} \mathrm{\; cm}$ have a lower limit on abundance of $\sim 10^{25} \mathrm{\; AU^{-3}}$ \citep{2000JGR...10510303L} (since this value represents the density inside of the heliosphere), and rogue planets with a radius of $\sim 10^{9} \mathrm{\; cm}$ may number $\sim 2$ per star \citep{2020AJ....160..123J}, which would correspond to a local number density of $\sim 0.3 \mathrm{\; pc^{-3}}$. We use these abundances in concert with the possible range for Borisov-like objects to construct a range of possible broken power-law size distributions, although we constrain the range of interest to $10^{-1} - 10^8 \mathrm{\; cm}$ (shaded range in Figure \ref{fig:borisov_size_dists}) given that sub-mm meteoroids and planets can be compositionally distinct, since dust grains are subject to unique dynamical considerations. While a break in the size distribution of interstellar objects does not necessarily exist at the size of Borisov, we adopt Borisov as a fiducial break point given the fact that the Poisson uncertainty associated with extrapolating its detection to a true abundance dominates the range of possible power-law fits. Figure \ref{fig:borisov_size_dists} indicates that an unbroken power-law distribution with a slope $q = 3$ is consistent with observations, representing a scale-free size distribution in which equal mass is contained in each logarithmic bin of size.

Based on the composition of its coma, Borisov was likely $\sim 97 \%$ carbon and oxygen by mass \citep{2020NatAs...4..867B}. Adopting this fraction and a typical mass density of $0.7 \mathrm{g \; cm^{-3}}$ \citep{2008M&PS...43.1033W}, we integrate over the size distributions displayed in Figure \ref{fig:borisov_size_dists} for the total mass of carbon and oxygen locked in interstellar objects. We compare the results to the implied mass of carbon and oxygen contained in nearby stars and the interstellar medium (ISM), given respective total densities $\rho_{\star} \simeq 0.036 \mathrm{\; M_{\odot} \; pc^{-3}}$ and $n_H \simeq 1.1754 \mathrm{\; cm^{-3}}$ \citep{2015ApJ...814...13M}, and applying Solar photospheric composition to both stars and the ISM \citep{2009ARA&A..47..481A}.

\begin{figure}
 \centering
\includegraphics[width=1\linewidth]{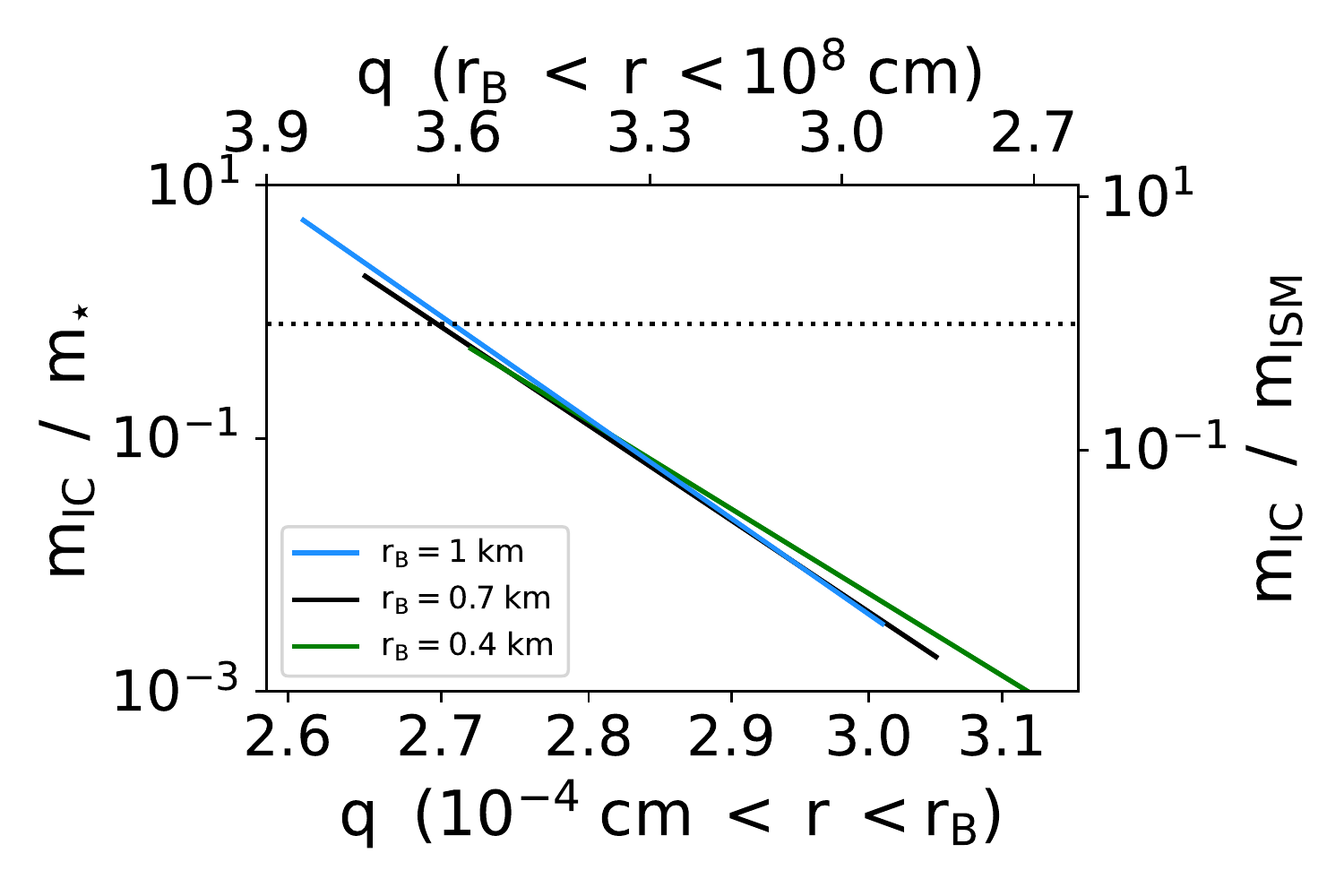}

\caption{Implied mass fraction of carbon and oxygen contained in interstellar objects, relative to stars (left y-axis) and the ISM (right y-axis) as a function of the value of $q$, the power-law index, for the lower (bottom x-axis) and upper (top x-axis) legs of the interstellar size distribution, which are split at the inferred radius of Borisov, $r_B$. The three solid lines indicate three different assumptions for the radius of Borisov, and the dotted line indicates the condition, $(m_{IC} / m_{ISM}) = 1$.}
\label{fig:slopes_vs_fractions}
\end{figure}

Figure \ref{fig:slopes_vs_fractions} shows the mass fraction of carbon and oxygen contained in interstellar objects relative to stars and to the ISM, under various asssumptions about the size distribution of interstellar objects and the radius of Borisov's nucleus. The results indicate that the fraction of carbon and oxygen contained in interstellar objects is sensitive to the size distribution of interstellar objects, varying by nearly four orders of magnitude depending on the parameters chosen for the broken power-law size distribution consistent with the detection of Borisov. Bounds of $\pm 0.7$ about the scale invariant value of $q = 3$ are necessary for the mass of heavy elements (metals) contained in interstellar objects not to exceed that of stars or the ISM. A wide range of metals mass fractions are consistent with present constraints, including a major mass fraction of carbon and oxygen contained in interstellar objects for a shallower power-law slope on smaller scales and steeper power-law slope and larger scales. The range of allowable size distribution slopes that fit the discovery of Borisov dominate the error budget relative to uncertainty associated with the abundance of rogue planets. However, if Borisov-like objects are discovered in the future and the uncertainty regarding the abundance of comet-sized interstellar objects is reduced, then the amount of metals locked in ISOs will depend more sensitively on the abundance of rogue planets.

Interestingly, the conservative scale-free power-law fit of $q = 3$ implies that nearly $\sim 1 \%$ of the carbon and oxygen contained in stars and the ISM is locked in interstellar objects. The minimum mass Solar nebula model requires $\sim 1\%$ of the Sun's mass to form the planets \citep{2007ApJ...671..878D}, so the result derived here implies a metals budget for interstellar objects comparable to the metals budget of the minimum mass Solar nebula model, implying that if interstellar objects are formed in protoplanetary disks a significant proportion of protoplanetary material is ejected during the planetary formation process. Additionally, if some interstellar objects have a finite lifetime, a larger mass budget is required. This presents a major challenge for recent planetary system formation simulations, which result in an ejected object mass density several orders of magnitude too low to be consistent with this constraint \citep{2020arXiv201108257P}. This conclusion is consistent with the difficulty in reconciling the implied abundance of interstellar objects with planetary system origins \citep{2018ApJ...866..131M, 2019AJ....157...86M}. Finally, we note that these results are not sensitive to the boundary conditions adopted here. Adjusting the assumed interstellar planet and dust densities, which are both uncertain, by an order of magnitude, results in only $\sim 20 \%$ and $\sim 30 \%$ changes in $m_{IC}$, respectively, and thereby does not affect significantly the results.

\section*{Acknowledgements}
This work was supported in part by a grant from the Breakthrough Prize Foundation.

\section*{Data Availability}
No new data were generated or analysed in support of this research.




\bibliographystyle{mnras}
\bibliography{example} 





\bsp	
\label{lastpage}
\end{document}